\documentclass[onecolumn,nofootinbib,colorlinks,hyperindex,10.5pt]{revtex4-2}
\usepackage{graphicx}
\usepackage{graphicx}
\usepackage{tikz}
\usepackage{booktabs}
\usepackage{bbding}
\usepackage{float}
\usepackage{simpler-wick}
\usepackage{pifont}

\usepackage{adjustbox}
\usepackage[compat=1.1.0]{tikz-feynman}
\usepackage{verbatim}
\usepackage{bbm}
\usepackage{amsfonts,amsmath,amssymb,tikz,accents}
\usepackage[eulergreek]{sansmath}
\usepackage{indentfirst,bigints}
\usepackage{hyperref,upgreek}
\usepackage{bm,graphics,color}
\usepackage{feynmp-auto}
\usepackage{slashed}
\counterwithout{equation}{section}
\usepackage{hyperref}
\usepackage{pgfplots}
\pgfplotsset{compat=1.16}
\usepackage{dsfont}
\hypersetup{hidelinks,backref=true,pagebackref=true,hyperindex=true,colorlinks=true,breaklinks=true,urlcolor= blue}
\hypersetup{%
  colorlinks = true,
  linkcolor  = blue,
  citecolor = cyan,
}
\usepackage{textgreek}
\usepackage{color,xcolor}
\newcommand{\gdualn}[1]{\overset{\:{}^{{}^{\boldsymbol{\neg}}}}{\smash[t]{#1}}} %ELKO dual

\def\0{\mbox{\boldmath$\displaystyle\mathbb{O}$}}
\usepackage{bbold}
\def\I{\openone}
\def\openone{\mathbb I}

\def\p{\mbox{\boldmath$\displaystyle\boldsymbol{p}$}}

\newcommand{\orcidicon}{%
	\begin{tikzpicture}
	\draw[lime, fill=lime] (0,0)
		circle [radius=0.16]
		node[white] {{\fontfamily{qag}\selectfont \tiny ID}};
	\draw[white, fill=white] (-0.0625,0.095)
		circle [radius=0.007];
	\end{tikzpicture}	\hspace{-2mm}
}
\newcommand\orcidRR{{\href{https://orcid.org/0000-0002-8283-2577}{\orcidicon}}}
\newcommand\orcidg{{\href{https://orcid.org/0000-0002-7942-7941}{\orcidicon}}}

\usepackage{float,xcolor,upgreek}
\usepackage{color} 
\usepackage{tikz-feynman}
\tikzfeynmanset{compat=1.0.0}
\newcommand{\beq}{\begin{eqnarray}}
\newcommand{\eeq}{\end{eqnarray}}
\newcommand{\bea}{\begin{eqnarray}}
\newcommand{\eea}{\end{eqnarray}}

\usepackage{bm}
\begin{document}

\title{Spinor Adjoints, Gauge Invariance and a new Road to Sterile Neutrinos}

%%%%%%%%%%%%%%%%% SUGESTÕES PARA O TÍTULO %%%%%%%%%%%%%%%%% 
%%%  The Interplay of Local Gauge Invariance and Discrete Symmetries
%%% Discrete Symmetries in the Framework of Local Gauge Theories
%%% Discrete Symmetry Constraints in Local Gauge Theories 
%%% Characterizing Local Gauge Transformations through Discrete Symmetries

\author{R. J. Bueno Rogerio\orcidRR{}}
\affiliation{Centro Universitário UNIFAAT, Atibaia-SP, 12954-070, Brazil.}
\email{rodolforogerio@gmail.com}

\author{G. B. de Gracia\orcidg{}}
\affiliation{Departamento de Física, Universidade Federal do Triângulo Mineiro UFTM,\\
38.025-180, Uberaba, MG, Brazil}
\email{gabriel.gracia@uftm.edu.br}

%%%%%% Abstract %%%%%%
\begin{abstract}

\indent In this work, we analyze the possibilities of certain gauge transformations regarding some specific spinorial dual structures. To this end, we define a general structure, which can be expressed in terms of discrete symmetry operators associated with parity, charge conjugation, and time reversal. To this end, we consider the demand of tensor covariance of the spinor bilinears under space-time transformations and also deep algebraic considerations such as the invertibility of the dual structure which guarantee that the physical information remains unchanged (respecting the Fierz-Pauli-Kofink identities), and so on. Later, we examine the global gauge freedom implied by each choice of dual and point out the physical implications. The final goal is achieved by establishing a well-defined coupling between the Elko and standard model neutrinos in a prelude for a new formulation of the hypothetical so-called sterile neutrino.
\end{abstract}

\maketitle

%%%%%% Main Text %%%%%%

\section{Opening Section}

In the context of spinor theory, we can say that the most prominent representatives are Dirac, Majorana, and Weyl spinors. Dirac spinors play a fundamental role in describing the electron, whereas Weyl spinors are associated with massless fermions. Regarding Majorana spinors, they are related to hypothetical descriptions of dark matter and sterile neutrinos, for example. All these spinors belong to the Lounesto classification, which considers geometric features among bilinear forms (physical information) to characterize spinors, categorizing them into six distinct classes \cite{lounestolivro}. However, as we know, physics is constantly evolving, and new discoveries are expected to be made. In this way, we state that the realm of spinors is likely not complete with only the cases mentioned above, and new spinors classes may emerge, bringing new directions and possible phenomenological descriptions within the quantum field theory framework.

That said, we can elucidate a theoretical discovery that has recently attracted attention. The last two decades have been highly significant for the development of the quantum field theory, particularly in the context of the spinor theory. This is due to the theoretical discovery of Elko spinors, a German acronym for eigenspinors of the charge conjugation operator. These spinors are also known as dark spinors, because they do not carry charge, cannot interact electromagnetically, only gravitationally \cite{jcap, mdobook}. It is worth mentioning that Elko occupies a new class in the Lounesto classification, being essentially distinct from the previously mentioned types of spinors.\\
The research on the properties of the Elko spinors shed light on new mathematical and physical concepts of QFT, since they present different characteristics if compared to Dirac spinors, for instance. We can highlight Elko's key characteristics: a more involved dual structure, wave equation (that is not first-order in momentum), quantum field propagator's structure, Hamiltonian operator, \emph{two-fold degeneracy feature} (Wigner degeneracy), among others, which strongly distinguish them from usual spinors \cite{beyondlounesto, rodolfonogo,dharamnogo,dharambosons,dharamspinstatistic,elkopolar,chenggeneral,chenglagrangian,juliodoublets}.

The characteristics mentioned above, combined with the mass dimensionality of the associated quantum field \cite{jcap, jhepgabriel}, made Elko spinors the focus of numerous studies in the context of cosmology and phenomenology, bringing to light promising results related to dark matter description \cite{saulo1, saulo2,saulo3,saulo4,cylcosmelkology,roldao2023,roldaofermionic,julioperturbative,chengyukawa,elkoherm}.

More specifically, these spinors are interesting candidates to describe massive particles whose associated quantum fields define irreducible representations of the extended Poincaré group \cite{jhepgabriel} being a well-defined dark matter candidate. Since the mass dimensionality of this spinor field  \cite{aaca, mdobook} imposes a limited possibility of coupling with gauge fields, and also avoids participation in standard model doublets,  it reinforces the possible interpretation in terms of dark matter. In summary, Elko spinors differ significantly from Dirac spinors with respect to a variety of properties \cite{jcap}, including the global symmetries, the focus of the present discussion. We highlight that Elko spinors and Dirac spinors exhibit intrinsically different possibilities for interaction through local gauge fields. While Dirac spinors support $U(1)$ gauge interactions, Elko spinors do not. Although this symmetry is forbidden by the charge conjugacy requirement, in \cite{dharamnpb}, one defines a quantum field based on these spinors as expansion coefficient taking into account the so-called Wigner degeneracy. Then, considering the whole quantum field, this requirement can be relaxed. Interestingly, according to \cite{elkoherm}, this symmetry cannot be made local for theories involving this field,  avoiding interactions such as the electromagnetic one. Thus, it becomes evident that different spinors classes in Lounesto sense carry different characteristics, both from a physical and mathematical perspective. This leads us to explore, in a general sense, how these particularities are associated with each type of spinor.

Although the traditional approach rooted in the Dirac dual structure has served as a cornerstone of modern high-energy physics, the advent of beyond-the-standard-model theories invites a reevaluation of this foundation. The incorporation of novel adjoint structures could transcend mere mathematical formalism \cite{mdobook, beyondlounesto, rjdual}, potentially unlocking profound insights into significant theoretical and physical extensions.Therefore, within the extensive spectrum of potential dual structures, it is essential to meticulously examine the relevant insights that they may offer. In the context of physical theories beyond the standard model, grounding the exploration of novel possibilities on a robust mathematical framework emerges as a judicious and methodical approach, laying the foundation for further advancements in theoretical physics. Following this route unraveled the whole class of singular spinors \cite{rrsingular2024}, including their most important representatives in the context of dark matter physics \cite{dharamnpb, chengtipo4,elkostates, jhepgabriel, dharamreports} the Elko ones, eigenstates of the charge conjugation operator. \\
 As mentioned,	\cite{elkoherm} shows that imposing the demand of Hermiticity and charge conjugation invariance for the whole action describing the interaction of mass dimension one fermionic  fields based on Elko spinor expansion coefficients \cite{dharamnpb} with gauge fields, one concludes that the minimal coupling procedure cannot be implemented. More specifically, one can prove that the dark nature of this fermion field arises from its defining adjoint structure and the restrictions from the most fundamental requirements of quantum field theory. Therefore, it motivates us to focus our study on global symmetries without necessarily referring to a specific form of a given Lagrangian.    

The aim of the study conducted in this work is, in light of all the peculiarities carried by a variety of spinors types \footnote{Obviously including the Elko ones.}, to perform a thorough analysis regarding the freedom associated with global gauge transformations, focusing our attention on how their specific dual structure allows some of these symmetries. It is also important to remember that the dual structure plays a fundamental role in spinor theory, since it carries a significant part of the physical information. For this purpose, we define a wide set of dual structures in terms of discrete symmetry operators and all their possible combinations, according to earlier works such as \cite{rjdual, rodolfohidden}. In principle, we develop a general discussion taking into account the properties of a set of spinor dual formulations and also considering various kinds of spinors, including the standard Dirac ones. \\
\indent For the specific case of Elko, the fact that these spinors are eigenstates of the charge-conjugation operator is also crucial to restrict the allowed types of global symmetries addressed here.
Therefore, we show that Elko spinors, due to their specific dual structure, are compatible with a chiral symmetry even for the massive case. This is a remarkable property with deep physical implications/applications.

This indicates that Elko spinors are suitable candidates for the so-called sterile neutrino \cite{sterile}, whose Lagrangian can be split into two independent chiral contributions \cite{grf}. Such an important outcome is another fundamental difference arising from the correct choice of its adjoint structure and correlated symmetry transformations.  In order to guarantee physically and mathematically well-founded results, we look towards ensuring the invariance/covariance of the dual structures under a general global transformation, and also ensure that such transformations do not affect the physical information carried by the spinors. To this end, we check if the bilinear forms are preserved and, additionally,  verify whether the algebraic relations between such bilinear forms comply with the geometric constraint known as the Fierz-Pauli-Kofink identities \cite{lounestolivro}.

The main results and novelties that can be highlighted are, firstly, related to the algebraic aspects. We associated the allowed range of the global symmetries of a spinor theory with the specific choice of the adjoint structures. This choice has direct physical implications, such as: invariance of the norm and bilinear forms, and specific constraints to ensure that a given spinor remains an eigenspinor of the charge conjugation operator (Elko) and parity (Dirac).  These general algebraic considerations form the basis for establishing a chiral-preserving effective quadratic coupling between fields based on Elko spinors with the neutrinos. Namely, this interaction defines a suitable mass-generating mechanism for the latter particle.

The paper is organized as follows: in the next section, we analyze the compatibility of the dual structures with the demand of covariance of spinor bilinears under space-time transformations. In Sect.\ref{dualandgauge} we investigate the main properties of global gauge transformations compatible with the set of allowed well-defined classes of spinor duals, looking towards the invariance of the dual structure and the verification of the Fierz-Pauli-Kofink (FPK) identities. We reserved Sect.\ref{transformations} to explore the properties of the allowed symmetry transformations according to a specific dual structure and kind of spinor field. The main result of this section, and also the central achievement of the paper is discussed at the end. It contemplates the recent results on the quantum field constructed upon Elko spinors \cite{mdobook} as well as the role that the chiral symmetry can play in an alternative discussion of neutrino oscillations \cite{bilenky}. Finally, in Sect.\ref{conclusion} we conclude.

\section{Space-time transformation properties}\label{dualtransformation}

In this section, we review the main aspects of the space-time transformation for a given spinor and its associated dual, taking forward previous preliminary investigations \cite{rjdual}.  
Let $S(\Lambda)$ denote the representation of a general Lorentz transformation  for (bi-) spinors whereas the transformed objects are denoted as follows $\uplambda^{\prime}(p^{\mu})=S(\Lambda)\uplambda(\Lambda^{-1}p^\mu)$. We consider the following structure for the adjoint as $\stackrel{\neg}{\uplambda}(p^\mu)=[\Delta \uplambda(p^{\mu})]^\dagger \gamma_0$, where the explicit form of the orthochronous proper Lorentz group, $L_+^\uparrow$, transformation reads
\begin{equation}
S=e^{\frac{i}{4}\delta\omega^{\alpha\beta}\sigma_{\alpha\beta}},
\end{equation}
in which $\delta\omega^{\alpha\beta}$ stands for the symmetry parameters and $\sigma_{\alpha\beta}=\frac{i}{2}[\gamma_{\alpha},\gamma_{\beta}]$, denoting the generators of the transformations. The boost and rotation generators can be written, respectively, as follows
\begin{eqnarray}
\sigma_{0i} = -\frac{i}{2}\left(\begin{array}{cc}
\sigma_{i} & 0 \\ 
0 & -\sigma_{i}
\end{array}\right) \quad \mbox{and}\quad \sigma_{ij} = \frac{1}{2}\epsilon_{ijk}\left(\begin{array}{cc}
\sigma_{k} & 0 \\ 
0 & \sigma_{k}
\end{array}\right).   
\end{eqnarray}
In which $\sigma_i$ stands for the Pauli matrices. Here, $\Delta$ can be represented by the discrete symmetry operators: where the covariant parity operator which acts on the functional form as well as on the spinor argument reads\footnote{As interestingly observed in \cite{speranca}, fundamental for all papers regarding Elko, the Dirac operator encodes the whole parity transformation regarding the action on the functional form as well as in the argument. } $\mathcal{P}=m^{-1}\gamma_{\mu}p^{\mu}$ \cite{speranca}, charge-conjugation $\mathcal{C}=\gamma_2 \mathcal{K}$ --- where $\mathcal{K}$ complex
conjugate to its right--- time-reversal $\mathcal{T}=i\gamma_5\mathcal{C}$ \cite{jcap} as well as its combinations. 
%By using the standard relation $S^{-1}=\gamma_0S^\dagger\gamma_0$,
The transformed candidates of scalar bilinears can be expressed as 
\begin{equation}      	
	\stackrel{\neg}{\uplambda^{\prime}}(p^{\mu})\uplambda^{\prime}(p^{\mu})=\uplambda{^\dagger}(\Lambda^{-1}p^{\mu}) S^{\dagger}\Delta^{\prime\dagger}(S^{-1})^\dagger\gamma_0\uplambda(\Lambda^{-1}p^{\mu}),\label{l4}
\end{equation} 
since $\Delta$ can depend on momentum, if one takes into account the combinations involving the parity operator introduced above, the $\Delta$ may also present a non-trivial transformation, thus, $\Delta'$ stands for the transformed $\Delta$. If $[\Delta,S]=0$, it is a sufficient condition to ensure a scalar behavior for the spinor norm. Nonetheless, if such a condition is not accomplished, the spinor norm composed by a given dual is covariant if its transformation is given by $\Delta'=S\Delta S^{-1}$. In fact, starting from the $\Delta$ operator transformation, it can be readily seen that $\Delta^\dagger=S^\dagger\Delta'^\dagger (S^{-1})^\dagger$. By plugging the last relation into \eqref{l4}, one concludes that the covariance is guaranteed. Accordingly, this requirement translates into 
\begin{equation}
[S^\dagger\Delta'^\dagger(S^{-1})^\dagger-\Delta^\dagger]=0.\label{l6}
\end{equation} 
being equivalently expressed as $\Delta^\dagger=S^\dagger\Delta'^\dagger (S^{-1})^\dagger$ and automatically $\Delta'=S\Delta S^{-1}$. We remark that with this condition, the transformed dual reads $\stackrel{\neg}{\uplambda'}(p_\mu)=\uplambda^\dagger(p_\mu)\Delta^\dagger\gamma_0S^{-1}=\stackrel{\neg}{\uplambda}(p_\mu) S^{-1}$. According to \cite{rjdual} and \cite{epjc2016}, the transformations outlined above will render the right covariance of the bilinears for a certain specific set of $\Delta$. For these specific cases, the bilinear forms will keep their characteristics under Lorentz transformations, for instance, the scalar ones will transform as a scalar, the ones presenting a vector index will transform as a vector, and so on. 

Here, we summarize some important results in light of \cite{rjdual}, checking the covariance and the additional algebraic constraint expressed as $\Delta^{\dag}\gamma_0=\gamma_0\Delta$ \cite{juliorogpla, juliorogepjc}. Such a constraint evinces an invertible adjoint structure in compliance with the Lorentz covariance. 

Thus, the covariance constraint for the $\Delta$ operator can be listed below
\begin{table}[H]
\centering
\begin{tabular}{c|cc}
\hline 
\;\;\;\;$\Delta$\;\;\;\; & \;\;\;\;$[\Delta,S]=0$\;\;\;\; &\;\;\;\; $\Delta^{\dag}\gamma_0=\gamma_0\Delta$\;\;\;\; \\ 
 \hline
$\I$ & $\checkmark$ & $\checkmark$ \\ 

$\mathcal{C}$ &  $\text{\sffamily X}$ &  $\text{\sffamily X}$ \\ 
 
$\mathcal{P}$ & $\checkmark$ & $\checkmark$ \\ 
 
$\mathcal{T}$ & $\text{\sffamily X}$ & $\text{\sffamily X}$ \\ 

$\mathcal{CP}$ & $\text{\sffamily X}$ & $\checkmark$ \\ 
 
$\mathcal{CT}$ & $\checkmark$ & $\checkmark$ \\ 
 
$\mathcal{PT}$ & $\text{\sffamily X}$ & $\text{\sffamily X}$ \\ 

$\mathcal{CPT}$ & $\checkmark$  & $\checkmark$ \\ 
 \hline 
\end{tabular} 
\caption{Results for the Lorentz covariance and algebraic constraint for each adjoint definition.}
\end{table}
From the above results, we observe that duals defined upon $\Delta = \I$, $\Delta = \mathcal{P}$, $\Delta = \mathcal{CT}$ and $\Delta = \mathcal{CPT}$ are mathematically and physically interesting. Nonetheless, the combinations $\Delta = \mathcal{CT}$ and $\Delta = \mathcal{CPT}$, present significant challenges, as previously noted in \cite{rjdual} and \cite{rodolfohidden}. Firstly, since it is not invariant under the standard proper orthochronous Lorentz subgroup, and second, because the dual would change its sign under improper Lorentz transformations (related to discrete symmetries). Thus, considering this adjoint structure, the bilinear covariants would interchange positions: pseudo scalar and current would transform without the standard additional sign whereas the usual scalar, vector, and bivectors would become the pseudo-tensor quantities.

\section{Some global gauge transformation properties}\label{dualandgauge}

In this section, we explore some properties of symmetry transformations that act just on the functional form of the spinors, looking towards understanding how it is affected by the dual structure. In most textbooks, this type of transformation is well-known and defined for Dirac spinors; however, we want to explore what happens when we evade the standard Dirac dual structure prescription.

The establishment of a mathematically well-defined adjoint structure is of fundamental importance for a spinorial theory. As seen in \cite{grf}, improperly defined dual leads to problematic issues on locality, incompatibility of quantum field operators, a Hamiltonian with negative energy, and impossibility of particle interpretation, among other serious problems.

Then, in order to investigate the interplay between dual spinorial prescription and the associated allowed symmetries, we start by setting the following general transformation 
\begin{eqnarray}\label{gauge1}
\uplambda(p^\mu) \rightarrow \uplambda^{\prime}(p^{\mu}) = e^{i\mathfrak{a}\alpha}\uplambda(p^{\mu}),
\end{eqnarray}
being $\alpha\in \mathds{R}$ a small transformation parameter and $\mathfrak{a}$ standing for a $4\times 4$ unitary matrix. The adjoint transformed structure reads  
\begin{eqnarray}\label{gaugedual}
\stackrel{\neg}{\uplambda^{\prime}}(p^{\mu})= [\Delta^{\prime}\uplambda^{\prime}(p^{\mu})]^{\dag}\gamma_0,
\end{eqnarray}
combining $\eqref{gauge1}$ and $\eqref{gaugedual}$, it yields 
\begin{eqnarray}
\stackrel{\neg}{\uplambda^{\prime}}(p^{\mu}) = \uplambda^{\prime\dag}(p^{\mu})e^{-i\mathfrak{a}^{\dag}\alpha}\Delta^{\prime\dag}\gamma_0.
\end{eqnarray}
Now, the requirement of an invariant norm leads to the following condition
\begin{eqnarray}
\stackrel{\neg}{\uplambda^{\prime}}(p^{\mu})\uplambda^{\prime}(p^{\mu})-\stackrel{\neg}{\uplambda}(p^{\mu})\uplambda(p^{\mu})=0,
\end{eqnarray} 
since we are focusing on global transformations at fixed points. Then, one gets
$[e^{-i\mathfrak{a}^{\dag}\alpha}\Delta^{\prime\dag}\gamma_0 e^{i\mathfrak{a}\alpha}-\Delta^{\dag}\gamma_0] =0$.
Rearranging the core of the last equation above, we reach the following condition
\begin{eqnarray}\label{gaugecond}
\Delta^{\prime}= \gamma_0 e^{i\mathfrak{a}^{\dag}\alpha}\gamma_0 \Delta e^{-i\mathfrak{a}\alpha},
\end{eqnarray}
such a result can be cast as a transformation of $\Delta$. Supposing a small transformation, \emph{i.e.} $|\alpha|<<1$, thus, equation \eqref{gaugecond} reads
   \begin{eqnarray}\label{gaugecondsmall}
\Delta^{\prime}=  \Delta - \Delta i\mathfrak{a}\alpha +\gamma_0 i \mathfrak{a}^{\dag}\gamma_0 \Delta\alpha.     
\end{eqnarray} 
If $\Delta^{\prime} = \Delta$, then the covariance of the bilinears is achieved; otherwise, we must check all the conditions on $\Delta$ and $\mathfrak{a}$ that lead to an invariant relation.

\section{Gauge Transformations in the adjoint context}\label{transformations}

Throughout this section, we list all the possible transformations mentioned above considering a set of specific dual structures. 
The main results are summarized as follows
\begin{table}[H]\label{tab2}
\centering
\begin{tabular}{c|cccccccc}
\hline 
\;\;\;\;$\Delta$\;\;\;\; & $\mathfrak{a}=\I$& \;\; $\mathfrak{a}=\gamma_0$\;\; & \;\;$\mathfrak{a}=\gamma_1$\;\;&\;\; $\mathfrak{a}=\gamma_2$\;\; &\;\; $\mathfrak{a}=\gamma_3$\;\; &\; $\mathfrak{a}=\gamma_5$ \;&\; $\mathfrak{a}=\gamma_{5}\gamma_{\mu}$\;&\;\;\;\; $\mathfrak{a}=\gamma_{\mu}\gamma_{\nu}$ with $\mu\neq\nu$\;\;\; \\ 
\hline
 $\I$ & $\checkmark$ (\emph{Dirac})  &$\checkmark$ &$\checkmark$ &$\checkmark$ &$\checkmark$ & $\text{\sffamily X}$& $\checkmark$&$\text{\sffamily X}$ \\

$\mathcal{C}$ &$\text{\sffamily X}$  & $\checkmark$ &$\checkmark$ & $\checkmark$ &$\checkmark$ & $\text{\sffamily X}$ &$\text{\sffamily X}$& $\checkmark$ \\ 
 
$\mathcal{P}$ & $\checkmark$ (\emph{Dirac}) & $\text{\sffamily X}$ & $\text{\sffamily X}$ &$\text{\sffamily X}$ &  $\text{\sffamily X}$& $\checkmark$ (\emph{Elko})&$\text{\sffamily X}$&$\text{\sffamily X}$\\ 
 
$\mathcal{T}$ & $\text{\sffamily X}$  & $\text{\sffamily X}$ & $\text{\sffamily X}$ &$\text{\sffamily X}$ &$\text{\sffamily X}$ & $\text{\sffamily X}$&  $\checkmark$& $\checkmark$ \\ 
 
$\mathcal{CP}$ & $\text{\sffamily X}$  & $\text{\sffamily X}$ &$\text{\sffamily X}$ & $\text{\sffamily X}$&$\text{\sffamily X}$ & $\checkmark$ &$\text{\sffamily X}$&$\text{\sffamily X}$ \\ 
 
$\mathcal{CT}$ & $\checkmark$ & $\text{\sffamily X}$ &$\text{\sffamily X}$ &$\text{\sffamily X}$ &$\text{\sffamily X}$ &$\text{\sffamily X}$ &$\text{\sffamily X}$&$\text{\sffamily X}$\\ 
   
$\mathcal{PT}$ & $\text{\sffamily X}$  &$\text{\sffamily X}$ &$\text{\sffamily X}$ &$\text{\sffamily X}$ &$\text{\sffamily X}$ & $\checkmark$&$\text{\sffamily X}$ &$\text{\sffamily X}$\\ 

$\mathcal{CPT}$ & $\checkmark$ & $\text{\sffamily X}$  & $\text{\sffamily X}$ &$\text{\sffamily X}$  &$\text{\sffamily X}$  & $\checkmark$ &$\text{\sffamily X}$&$\text{\sffamily X}$\\ 
\hline 
 \end{tabular} 
 \caption{All the possibilities of gauge transformations according to a specific dual structure prescription. In the last two columns of the table, the symbol $\checkmark$ indicates that there is \emph{at least} one combination that fulfills the desired condition.}
\end{table}

We highlight a key result concerning the previous conclusions and Elko spinors: for the transformed spinor to preserve charge conjugacy, we have the following constraints 
\begin{eqnarray}
\mathcal{C}\uplambda^{\prime}&=&\pm \uplambda^{\prime}, \label{Ctransformado}\\
&=&\pm e^{i\mathfrak{a}\alpha}\uplambda, \nonumber
\end{eqnarray}
verified if in equation \eqref{Ctransformado}, associated with the charge conjugation operator $\mathcal{C}=\gamma_2\mathcal{K}$, the following condition holds $\gamma_2\mathfrak{a}^{*} = -\mathfrak{a}\gamma_2$. While it is well known that the condition is fulfilled by any of the $\gamma_\mu$ matrices, for the case of Elko spinor and its adjoint, the only non-trivial solution also satisfying the fundamental demand of norm invariance \eqref{gaugecond} stands for $\mathfrak{a} = \gamma_5$. It is a remarkable result, ensuring a chiral symmetry suitable for defining a well-posed interaction with neutrinos.

A similar analysis must be carried out for Dirac spinors. The transformed spinor should still be an eigenspinor of the parity operator, leading to the following constraint: $[\mathcal{P}, \mathfrak{a}] = 0$. This relation only holds if $\mathfrak{a} = \I$, establishing the global U$(1)$ symmetry. The local version is achieved through considerations on the specific Lagrangian structure and the inclusion of gauge fields in the minimal coupling procedure.

An important point to emphasize is that the results presented in the table above are general, meaning they are independent of the specific type of spinor or its class within Lounesto's classification. The key aspect here is ensuring the right space-time covariance of the bilinear forms, while preserving the extra demand of either parity or charge conjugation invariance.

Regarding the bilinear densities relations $\rho_i=\stackrel{\neg}{\uplambda}(p^\mu)\Gamma_i\uplambda(p^\mu)$, where the homogeneous elements of the space-time Clifford algebra, $\Gamma_i$, are chosen accordingly \cite{lounestolivro, beyondlounesto}
\begin{equation}\label{cliffordbasis}
\Gamma = \lbrace \mathbbm{1}, \gamma_{\mu}, \gamma_{0123}, \gamma_{\mu}\gamma_{0123}, \gamma_{\mu}\gamma_{\nu}\rbrace, 
\end{equation}
remains unchanged, \emph{i.e.} $\rho^{\prime}_i=\rho_i$, and obeying the FPK identities, only for the relations ``$\checkmark$'' marked above. This is a new and important result. To have a physically acceptable theory, we must seek the duals and the respective transformations that preserve the information carried by the spinors unchanged.

However, the covariance of such bilinear forms depends not only on the data from Table II (related to fixed point symmetries), but also on the conditions specified in Table I. In addition, not all duals shown in Table I allow bilinear forms that satisfy certain geometrical conditions, namely FPK identities ---  quadratic relations that the bilinear covariants must obey --- as shown in \cite{rjdual, rodolfohidden}. Therefore, Lagrangian theories constructed upon the use of these exotic duals certainly present a more challenging physical interpretation. It can define an interesting topic for further investigations.

Connecting the results presented in the two tables, we observe that the identity matrix trivially satisfies the covariance conditions and the algebraic constraint. Although this transformation is associated with the global gauge freedom for the case of Dirac spinors (\emph{only}), it cannot be applied to the Elko spinors if one wants to keep their behavior under charge conjugation. The only form that $\Delta$ can take, besides keeping Lorentz covariance and respecting the algebraic constraint, stands for $\Delta = \mathcal{P}$. Such an output evinces, once again, that the parity operator plays a fundamental role in spinor theory. For the specific case of the Dirac spinors, $\Delta = \mathcal{P}$ leads to the usual Dirac adjoint structure. Then, since Dirac spinors are eigenspinors of the parity operator, $\mathcal{P}\psi=\psi$, \cite{mdobook, rrdirac1}.\\
\indent The freedom associated with the specific dual structure compatible with Elko, a representative of a class of singular spinors \cite{jrold1, out}, allows the separation of the field, as well as the Lagrangian, in independent pieces defined by its chiral modes, even for the case of massive particles \cite{grf}. This is a remarkable result, with phenomenological implications, arising from the deep conceptual algebraic considerations. Moreover, it enables one to consider Elko as a good candidate for a sterile neutrino, a hypothetical particle that is also associated with a dark matter candidate.\\
\indent Therefore, following the philosophy of the recent achievement on the Hermitian formulation of field theories based on the Elko construction \cite{elkoherm}, one can consider the global freedom associated with the Elko dual to establish a Hermitian quadratic interaction with a given flavor of neutrino \footnote{$\bar \nu(x)$ denotes the dual of the neutrino field taken with the standard Dirac adjoint. We are highlighting the whole action and focusing here on only one flavor to just demonstrate the overall mechanism.  } $\nu(x)$, namely, just the left chiral ones according to the Wu experiment \cite{wu}. Although it refers to the existence of just left-handed helicity measurements, focusing on a left chiral field allows us to include mass to it through a suitable coupling in such a manner to relativize this conclusion ensuring, in this case, at least predominance of left-handed helicity, 
\begin{align}
 \mathcal{S} = \int d^4x\Bigg[\Big(&\partial_\mu \gdualn{\uplambda}_{\mathrm{R}}(x)\partial^\mu\uplambda_{\mathrm{R}}(x)-m^2\gdualn{\uplambda}_{\mathrm{R}}(x)\uplambda_{\mathrm{R}}(x) +\partial_\mu \gdualn{\uplambda}_{\mathrm{L}}(x)\partial^\mu\uplambda_{\mathrm{L}}(x)-m^2\gdualn{\uplambda}_{\mathrm{L}}(x)\uplambda_{\mathrm{L}}(x)\Big)\nonumber \\ &+i\bar{\nu}_L\slashed{\partial}\nu_L+ g'\big(\bar{\nu}_L i\slashed{\partial}\uplambda+m\gdualn{\uplambda}\nu_L\big)\Bigg]
\end{align}

The $L/R$ label denotes the left/right chiral part of the spinor. This structure is due to the global chiral symmetry ensured by the structure of the Elko dual. Here, the left chiral neutrino can be written in terms of a four dimensional spinor structure using the projection $\nu_L=(I-\gamma_5)\nu$. Since the coupling preserves chirality due to the adjoint structure of the mass dimension one field, just the left chiral Elko components enter the interaction.\\
The coupling is chiral-preserving in accordance with the properties of the Elko adjoint prescription. Then, just its left sector is interacting with the neutrino assumed to present this same chirality. This quadratic coupling, as is going to be proved in a forthcoming paper, can be obtained from a more fundamental interaction involving the Higgs doublet and also the leptonic doublets, in which the neutrinos appear, in order to form a scalar under SU$(2)_L$ weak symmetry that can be coupled to the Elko field in a trilinear interaction. This latter fact complies with the fact that mass dimension one fermions based on Elko spinors cannot participate in standard model doublets due to its intrinsic mass dimension, reinforcing its dark matter/sterile neutrino interpretation. Then, after spontaneous symmetry breaking, the quadratic piece highlighted above is one of the main contributions. Here, $m$ denotes the Elko mass. A subtlety above is the fact that $\uplambda(x)$ is now associated with the specific spinor defined by Elko, and, besides that, encodes the whole quantum field constructed upon Wigner degeneracy \cite{dharamnpb}. 
Therefore, according to the recent paper \cite{elkoherm}, the adjoint of the mass dimension one field associated with Elko spinor expansion coefficients can be expressed as  
\bea \gdualn{\uplambda}(x)=\left(i\frac{\slashed{\partial}}{m} \uplambda(x)\right)^\dagger \gamma_0      \label{adjo}     \eea
enabling one to directly verify that $\mathcal{L}^{\textit{int}}$ is indeed Hermitian, ensuring a consistent probabilistic interpretation for the interaction. \\
\indent According to \cite{dharamnpb}, the corresponding quantum free field operator, in full compatibility with the rotational constraint thanks to the consideration of the Wigner degeneracy, is described below \footnote{The basis spanned by the self-conjugate sector $\xi_h(\p)$ as well as the anti-self-conjugate part $\upchi_h(\vec p)$ is defined in \cite{dharamnpb}. }
\bea \uplambda(x)=\int \frac{d^3p}{(2\pi)^3\sqrt{2mE} }\left[ \left(\sum_{h=1}^{4}c_h(\p)\xi_h(\p)  \right)e^{-ip.x}+\left(\sum_{h=1}^{4}d^\dagger_h(\p)\upchi_h(\vec p)  \right)e^{ip.x}   \right],            \eea
with the index $h=1...4$ referring to a basis constructed upon all the Elko spinor types (see \cite{dharamnpb, rrsingular2024} for details) with well-chosen coefficients. This is obtained through the consideration of the anti-linear nature of the charge conjugation operator. The aforementioned adjoint explicitly reads \footnote{Using the definition $\mathcal{P}=\frac{\slashed{p} }{m}$.}
\bea \gdualn{\uplambda}(x)=\int \frac{d^3p}{(2\pi)^3\sqrt{2mE}}\left[ \left(\sum_{h=1}^{4}c^\dagger_h(\p)(\mathcal{P}\xi_h(\vec p))^\dagger  \right)e^{ip.x}+\left(\sum_{h=1}^{4}d_h(\p)(-\mathcal{P}\upchi_h(\p))^\dagger  \right)e^{-ip.x}   \right] \gamma_0  .         \eea
in accordance with \eqref{adjo}.\\
\indent Interestingly, for $g'\neq 0$, one can consider the equations of motion for the fermions as well as their adjoints 
\bea  0=i\slashed{\partial}\bar{\nu}_L+g'm\gdualn{\uplambda}_L                                     \eea 
\bea  0=-(\Box+m^2)\gdualn{\uplambda}_L-g'i\slashed{\partial}\bar{\nu}_L                                                    \eea
with the last two equations leading to $(\Box+m^2+{g'}^2m)\gdualn{\uplambda}_L=0$, whereas the other equation of motion below
\bea \slashed{\partial}\nu_L=-g' \slashed{\partial}\uplambda_L                                       \eea
implies in $\nu_L=-g'\uplambda_L+\tilde{\Phi}$, with $\tilde{\Phi}$ denoting a general massless contribution in the kernel of the Dirac operator. Then, considering the equation 
\bea  -(\Box+m^2)\uplambda_L=-g'm\nu_L                \eea
together with $(\Box+m^2+{g'}^2m)\gdualn{\uplambda}_L=0$ one can prove that the massless neutrino contribution in the kernel of the Dirac operator vanishes and both fermionic fields obey the equation
\bea (\Box+m^{'\ 2})\nu_L(x)=0             \eea
with $m^{'\ 2}=m^2+g^{'\ 2}m$, since $g'$ is dimensionful because fermions based on Elko have mass dimension one whereas the neutrino has mass dimension three halves. Since the phenomenology implies a very small range for the mass parameters, these sterile neutrinos may be accompanied by heavier decoupled Elko fermions ensuring a standard cold dark matter scenario.
Therefore, the theory describes a legitimate interaction that can furnish, in forthcoming achievements, a possible description of neutrino oscillations.\\
\indent Basically, this coupling can generate mass for the neutrino keeping its left chiral nature. Moreover, although the external particles are neutral, the field itself, is not. It is a combination of self and anti-self-conjugate sectors with oscillatory operator coefficients, meaning that a global $U(1)$ symmetry is also allowed \cite{elkoherm}.
Then, the coupling also presents Lepton number symmetry (associated with the freedom by global U$(1)$ transformation). The first property guarantee that, although massive, as neutrino oscillation data implies, the right helicity contribution would be highly suppressed, making it possible to fix parameters to agree with experiments of the Wu type. The second property is related to a symmetry that prevents neutrinoless beta decay, something in accordance with the current observed phenomenology. Then, the first and second properties represent improvements with relation to Dirac and Majorana formulations, respectively. Regarding the latter, there are authors claiming that Majorana mass term may vanish classically and also when considered as a $q$-number, see \cite{vadim}, reinforcing the pertinency of our alternative approach.

\section{Concluding remarks}\label{conclusion}

In this work, we performed a comprehensive analysis of global gauge transformations based on the dual prescription for a given spinor adjoint. To achieve this, we considered a general dual structure, which can be consistently expressed in terms of discrete symmetry operators, and we then examined for each discrete symmetry, and their possible combinations, which types of transformations are physically allowed and relevant.\\
We highlight that all our results are independent of the spinors structure, depending solely on the operator that composes the adjoint spinor. As a significant result, which agrees with various other results in the current literature, the only dual prescription that lies in the intersection of all the physical demands is the one associated with the generalized parity \cite{speranca} operator. It can be reduced to the case of the Standard dual structure when considering the Dirac spinor \cite{rodolforegular}. For the case of Elko and all the correlated singular spinors, this is the only choice capable of retaining the most important physical constraints \cite{grf}, revealing the fundamental role played by such adjoint prescription.\\
\indent Finally, after these fundamental achievements, we considered the Elko spinor case and derived a suitable Hermitian coupling with the neutrino. Therefore,  the nature of its quantum field and the allowed chiral symmetry associated with the specific dual structure, ensured a set of interesting properties enabling an interpretation as an entirely new sterile neutrino candidate. This formulation complies with a wider set of phenomenological symmetries demonstrating the relevance of our achievement.

\acknowledgements
We acknowledge Professor Julio Marny Hoff da Silva for the privilege of his review and for the insightful discussions and observations during the writing process. We would like to thank Cheng-Yang Lee for correspondence. 
RJBR thanks the generous hospitality offered by UNIFAAT and to Prof.  Renato Medina. G.B. de Gracia thanks FAPESP post-doctoral grant (2021/12126-5) as well as the hospitality offered by UFTM.


\begin{thebibliography}{40}

\bibitem{lounestolivro}
P. Lounesto, \textit{Clifford algebras and spinors}, Second Edition, Editorial Cambridge University Press, Cambridge (2002).

\bibitem{jcap}D.~V.~Ahluwalia and D.~Grumiller,
Spin half fermions with mass dimension one: Theory, phenomenology, and dark matter,
JCAP \textbf{07}, 012 (2005).


\bibitem{mdobook}
D. V. Ahluwalia, {\it Mass Dimension One Fermions} (Cambridge Monographs on Mathematical Physics). Cambridge University Press, Cambridge, (2019).

\bibitem{beyondlounesto}
C. H. Coronado Villalobos, R. J. Bueno Rogerio, A. R. Aguirre, et al., On the generalized spinor classification: beyond the Lounesto’s classification, Eur. Phys. J. C {\bf 80}, 228 (2020). 

\bibitem{rodolfonogo}
 R. J. Bueno Rogerio, J. M. Hoff da Silva, C. H. Coronado Villalobos, Regular spinors and fermionic fields, Physics Letters A {\bf 402}, 127368 (2021).

 \bibitem{dharamnogo}
D. V. Ahluwalia, Evading Weinberg's no-go theorem to construct mass dimension one fermions: Constructing darkness, Europhys. Lett. {\bf 118}, 6, 60001 (2017).


\bibitem{dharambosons} D. V. Ahluwalia, Spin-half bosons with mass dimension three half: towards a resolution of the cosmological constant problem, EPL {\bf 131}, 41001 (2020).

\bibitem{dharamspinstatistic} D. V. Ahluwalia and C.-Y. Lee, Spin-half bosons with mass dimension three-half: Evading the spin-statistics theorem, EPL {\bf 140}, 24001 (2022).

\bibitem{elkopolar} L. Fabbri, ELKO in polar form, The European Physical Journal Special Topics {\bf 229}, 2117 (2020).

\bibitem{chenggeneral} C.-Y. Lee, Spin-half mass dimension one fermions and their higher-spin generalizations, The European Physical Journal
Special Topics {\bf 229}, 2003 (2020).

\bibitem{chenglagrangian} C.-Y. Lee, A lagrangian for mass dimension one fermionic dark matter, Physics Letters B \textbf{760}, 164 (2016).

\bibitem{juliodoublets}
F. A. da Silva Barbosa, J. M. Hoff da Silva, Non-standard Wigner doublets, EPL {\bf 144}, 5, 54001 (2023).

\bibitem{saulo1} S. H. Pereira, Degeneracy pressure of mass dimension one fermionic fields and the dark matter halo of galaxies, International
Journal of Modern Physics D {\bf 31} (2022).

\bibitem{saulo2} S. H. Pereira, M. E. S. Alves, and T. M. Guimar\~{a}es, An unified cosmological evolution driven by a mass dimension one
fermionic field, The European Physical Journal C {\bf 79} (2019).

\bibitem{saulo3} S. H. Pereira and R. S. Costa, Partition function for a mass dimension one fermionic field and the dark matter halo of
galaxies, Modern Physics Letters A \textbf{34}, 1950126 (2019).

\bibitem{saulo4}
S. H. Pereira, Degeneracy pressure of mass dimension one fermionic fields and the dark matter halo of galaxies, IJMPD \textbf{31}, 07, 2250056 (2022).

\bibitem{cylcosmelkology}
Cheng-Yang Lee, Haomin Rao, Wenqi Yu, Siyi Zhou, Cosmelkology: Elko fermions in FLRW space-time, arXiv:2402.05623 [hep-th] (2024).


\bibitem{roldao2023} G. de Gracia, A. Nogueira, and R. da Rocha, Fermionic dark matter-photon quantum interaction: A mechanism for
darkness, Nuclear Physics B {\bf 992}, 116227 (2023).

\bibitem{roldaofermionic} G. B. de Gracia and R. da Rocha, Fermionic dark matter interaction with the photon and the proton in the quantum
field-theoretical approach and generalizations,  arXiv:2206.11989v1 [hep-th] (2022).


\bibitem{chengyukawa} M. Dias and C.-Y. Lee, Constraints on mass dimension one fermionic dark matter from the Yukawa interaction, Physical
Review D {\bf 94}, 065020 (2016).

\bibitem{elkoherm}
G. B. de Gracia, et.al. On Wigner Degeneracy in Elko theory: Hermiticity and Dark Matter Phenomenological Constraints, arXiv:2407.00126 [hep-ph] (2024).

\bibitem{julioperturbative} W. Carvalho, M. Dias, A. C. Lehum, and J. M. H. da Silva, Perturbative aspects of mass dimension one fermions nonminimally
coupled to electromagnetic field, Eur. phys. Lett. {\bf 143}, 6 (2023).

\bibitem{elkostates} 
J. M. Hoff da Silva, and R. J. Bueno Rogerio, {\it Massive spin-one-half one-particle states for the mass-dimension-one fermions.} EPL {\bf 128}, 11002 (2019).

\bibitem{jhepgabriel}
D. V. Ahluwalia, et.al. Irreducible representations of the inhomogeneous Lorentz group with two-fold Wigner degeneracy,  J. High Energ. Phys. {\bf 75} (2024).

\bibitem{dharamreports}
D. V. Ahluwalia et.al, \emph{Mass dimension one fermions: Constructing darkness}, 	Physics Reports {\bf 967}, 1-43 (2022).

\bibitem{aaca} D. V. Ahluwalia, {\it The theory of local mass dimension one fermions of spin one half}, Adv. Appl. Clifford Algebras {\bf 27}, 2247 (2017).

\bibitem{dharamnpb} D. V. Ahluwalia, J. M. Hoff da Silva, C.-Y. Lee, Mass dimension one fields with Wigner degeneracy: A theory of dark matter, Nuclear Physics B {\bf 987}, 116092 (2023).

\bibitem{rjdual} J. M Hoff da Silva, R. J. Bueno Rogerio and N. C. R. Quinquiolo. Spinorial discrete symmetries and adjoint structures. Phys. Let. A {\bf 452}, 128470 (2022).


\bibitem{rrsingular2024}
R. J. Bueno Rogerio, C. H. Coronado Villalobos, Singular spinors as expansion coefficients of local spin-half fermionic and bosonic fields: On the two-fold Wigner degeneracy, Physics Letters A {\bf 498}, 129348 (2024).


\bibitem{chengtipo4} Lee, C-Y. Fermionic degeneracy and non-local contributions in flag-dipole spinors and mass dimension one fermions. Eur. Phys. Jour. C, {\bf 81}, 1 (2021).



\bibitem{rodolfohidden} R. J. Bueno Rogerio, R. T. Cavalcanti, C. H. Coronado Villalobos, J. M. Hoff da Silva, A note on hidden classes in spinor classification, 	arXiv:2405.06675 [math-ph] (2024).

\bibitem{sterile} B. Dasgupta and J. Kopp, Sterile Neutrinos,  Phys. Rept. {\bf 928}, 1-63 (2021).

\bibitem{grf}
G. B. de Gracia, R. J. Bueno Rogerio, L. Fabbri, Unraveling the Physical Meaning Behind Elko's Dual structure, 	arXiv:2408.16196 [hep-th] (2024).

\bibitem{bilenky}
S.Bilenky, Neutrino oscillations: From a historical perspective
to the present status. Nuclear Physics B {\bf 908}, 2 (2016). 


\bibitem{speranca}
L. D. Sperança, An Identification of the Dirac Operator with the Parity Operator, \emph{Int. J. Mod. Phys.} \textbf{D 2}, 1444003 (2014).

\bibitem{epjc2016}
J. M. Hoff da Silva, C. H. Coronado Villalobos, R. J. Bueno Rogerio, E. Scatena, On the bilinear covariants associated to mass dimension one spinors, Eur. Phys. J. C {\bf 76}, 563 (2016).





\bibitem{juliorogpla}
J. M. Hoff da Silvaa, R. T. Cavalcanti, Further investigation of mass dimension one fermionic duals, 	Phys. Lett. A {\bf 383}, 15 (2019).

\bibitem{juliorogepjc}
R. T. Cavalcanti, J. M. Hoff da Silva, Unveiling Mapping Structures of Spinor Duals, Eur. Phys. J. C {\bf 80}, 325 (2020).





\bibitem{rrdirac1}
C. H. Coronado Villalobos, R. J. Bueno Rogerio, The connection between Dirac dynamic and parity symmetry, Europhysics Letters {\bf 116}, 60007 (2016).

\bibitem{jrold1}
J .M. Hoff da Silva, R. da Rocha,
Unfolding Physics from the Algebraic Classification of Spinor Fields, 
\textit{Phys. Lett. B} \textbf{718}, 1519 (2013).

\bibitem{out} 
J. M. Hoff da Silva, R. T. Cavalcanti, Revealing how different spinors can be: the Lounesto 
spinor classification, \textit{Mod.Phys.Lett.A} \textbf{32}, 1730032 (2017).

\bibitem{rodolforegular}
R. J. Bueno Rogerio, J. M. Hoff da Silva, C. H. Coronado Villalobos, Regular spinors and fermionic fields, 	Physics Letters A {\bf 402}, 127368 (2021).



\bibitem{wu}
C. S. Wu; \emph{et al}. Experimental 7-Test of Parity Conservation in Beta Decay. Phys. Rev. {\bf 105}, 1413 (1957).

\bibitem{vadim}
V. Monakhov, Majorana Mass Term of Majorana Spinors, Phys. Sci. Forum {\bf 7}, 1 (2023).

\bibitem{gomez}
J. J. Gómez-Cadenas, \emph{et al}. The search for neutrinoless double-beta decay,  Riv. Nuovo Cim. {\bf 46}, 619 (2023).


\bibitem{helo}
J.C. Helo, S. Kovalenko and I. Schmidt, Sterile neutrinos in lepton number and lepton flavor violating decays. Nuclear Physics B {\bf 853} 80 (2011).


\begin{comment}







\bibitem{wigner1} Wigner E P, Unitary representations of the inhomogeneous Lorentz group including reflections, 1964
Group Theoretical Concepts and Methods in Elementary Particle Physics (Lectures of the Istanbul
Summer School of Theoretical Physics (1962)) ed F G¨ursey (New York: Gordon and Breach) 

\bibitem{wigner2} Wigner E P, On unitary representations of the inhomogeneous Lorentz group, 1939 Ann. Math. 40 149
Wigner E P, 1989 Nucl. Phys. Proc. Suppl. 6 9 (Reprinted)



\bibitem{weinberg1} S. Weinberg, {\it The Quantum Theory of Fields}, Vol. I: Foundations, Cambridge University Press, New York, (1996).

\bibitem{ryder}
L. H. Ryder, \textit{Quantum Field Theory}, Second Edition, Editorial Cambridge University Press, New York (1996).

\bibitem{pamdirac}
P. A. M. Dirac Paul, A theory of electrons and protons, Proc. R. Soc. Lond. A, 126360–365 (1930).




\bibitem{rrdirac2}
 R. J. Bueno Rogerio, C. H. Coronado Villalobos, Non-standard Dirac adjoint spinor: The emergence of a new dual, Europhysics Letters {\bf 121}, 21001 (2018).

 

\bibitem{ramond}
P. Ramond, \textit{Field Theory: A modern primer}, Second Edition, Editorial Addison Wesley-Publishing, California (1989).


 


\bibitem{rrtaka} R. J. Bueno Rogerio, R. T. Cavalcanti, J. M. Hoff da Silva, et.al.,
Revisiting Takahashi's inversion theorem in discrete symmetry-based dual frameworks, 	Physics Letters A, {\bf 481}, 129028 (2023).


\bibitem{rjtau} R. J Bueno Rogerio and J. M. Hoff da Silva, The local vicinity of spin sums for certain mass-dimension-one spinors. EPL, {\bf 118}, 10003 (2017).
 









  








\bibitem{sterile} B. Dasgupta and J. Kopp, Sterile Neutrinos,  Phys. Rept. {\bf 928} (2021) 1-63 

 \bibitem{nakahashi}
N. Nakanishi, and I. Ojima, \emph{Covariant operator formalism of gauge theories and quantum gravity}, World Scientific Publishing Co. Pte. Ltd., Editorial  Utopia Press, Singapore (1990).
\end{comment}

\end{thebibliography}
\end{document}